\documentclass{article}
\usepackage{arxiv}

\usepackage[utf8]{inputenc} 
\usepackage[T1]{fontenc}    
\usepackage{hyperref}       
\usepackage{url}            
\usepackage{booktabs}       
\usepackage{amsfonts}       
\usepackage{nicefrac}       
\usepackage{microtype}      
\usepackage{graphicx}
\usepackage{authblk}
\usepackage[version=3]{mhchem} 
\usepackage{siunitx}
\usepackage{bm}

\title{Photoemission spectroscopy using virtual photons emitted by positron sticking: A complementary probe for top-layer surface electronic structures}%
\author[1,*]{Alexander J. Fairchild}
\author[1,*]{Varghese A. Chirayath}
\author[2,3]{Bernardo Barbiellini}
\author[1]{Randall W. Gladen}
\author[1]{Ali R. Koymen}
\author[1]{Alex H. Weiss}
\affil[1]{Department of Physics, University of Texas at Arlington, Arlington, TX, United States}
\affil[2]{School of Engineering Science, Lappeenranta-Lahti University of Technology, Lappeenranta, Finland}
\affil[3]{Physics Department, Northeastern University, Boston MA 02115, USA}
\affil[*]{alexander.fairchild@mavs.uta.edu, chirayat@uta.edu}

\begin{document}
\maketitle

\begin{abstract}
We present a spectroscopic method which utilizes virtual photons to selectively measure the electronic structure of the top-most atomic layer. These virtual photons are created when incident positrons transition from vacuum states to bound surface states on the sample surface and can transfer sufficient energy to excite electrons into the vacuum. The short interaction range of the virtual photons restricts the penetration depth to approximately the Thomas-Fermi screening length. Measurements and analysis of the kinetic energies of the emitted electrons made on a single-layer of graphene deposited on Cu and on the clean Cu substrate shows that the ejected electrons originate exclusively from the top-most atomic layer. Moreover, we find that the kinetic energies of the emitted electrons reflect the density of states at the surface. These results demonstrate that this technique will be a complementary tool to existing spectroscopic techniques in determining the electronic structure of 2D materials and fragile systems due to the absence of subsurface contributions and probe-induced surface damage.
\end{abstract}

Photoemission spectroscopy (PES) is a powerful technique that has found success in the study of the electronic structure of molecules, solids, and surfaces \cite{Hufner2003, Mo2017}. The surface selectivity of PES relies on the inelastic mean free path (IMFP) of the escaping photoelectrons. For low-energy photoelectrons ($<$ 10 eV) the IMFP can be many atomic layers \cite{Olga2020} resulting in subsurface contributions to the PES spectrum. For example, in PES studies of single-layer graphene (SLG) grown on Cu foils, Cu contributions to the SLG spectrum were observed \cite{Avila2013}. A technique that probes the top-most atomic layer electronic structure with zero contribution from sub-surfaces or the substrate would therefore be a valuable complement to existing methods. Development of such a technique would be relevant to the research that engineers surface electronic structure to attain favorable catalytic or device properties \cite{Song2021, Pereira2009}. Auger-mediated positron sticking (AMPS) offers such a top-most atomic layer sensitive spectroscopic technique of the electron structure of 2D materials and surfaces.  

\begin{figure*}[ht]
    \centering
   \includegraphics[width=0.8\textwidth, keepaspectratio]{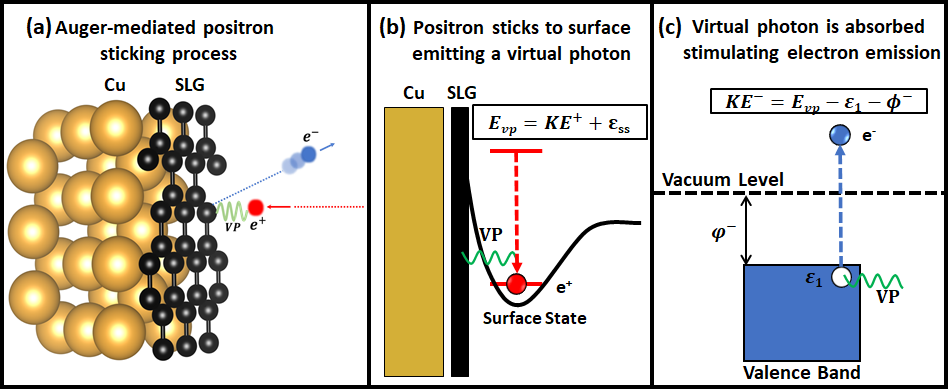}
    \caption{(a) The Auger-mediated positron sticking (AMPS) process illustrated for single-layer graphene (SLG) on Cu: an incoming low-energy positron (red) sticks to SLG transferring its energy, via a virtual photon (VP) depicted in green, to an electron (blue) which now has sufficient energy to escape the material. (b) The first step: the emission of a VP as a result of a low-energy positron making a transition from a vacuum state to a bound surface state. The VP energy is $E_{vp}=KE^+ + \varepsilon_{ss}$, where $KE^+$ is the positron kinetic energy and $\varepsilon_{ss}$ is the surface state binding energy. (c) The VP is absorbed by an electron in the valence band providing sufficient energy to liberate the electron from the material with kinetic energy $KE^- = KE^+ + \varepsilon_{ss} - \varepsilon_1 - \phi^-$, where $\varepsilon_1$ is the electron binding energy and $\phi^-$ is the electron work function.}
    \label{fig:amps}
\end{figure*}

Figure \ref{fig:amps} is a schematic of the AMPS process in which a virtual photon is emitted following the transition of a low-energy positron from a scattering state to an image-potential-induced surface bound state with sufficient energy to liberate an electron from the material. A virtual photon \cite{Nimtz2009,Stahlhofen2006} can excite electronic transitions like a ``real" photon; however, the key difference between a ''real" photon and a virtual photon is that the former penetrates deeply into the solid while the latter is screened rapidly penetrating only about an angstrom \cite{Appelbaum1975}. In particular, the virtual photon exchange of the AMPS interaction is spatially confined to within the Thomas-Fermi screening length of the surface \cite{Walker1992}. This ensures that AMPS is selective to only the top-most atomic layer of the solid. For example, in Cu, we estimate the Thomas-Fermi screening length to be $\sim$ 1 {\AA} \cite{Barbiellini1989}. 

Here, we present measurements of kinetic energy distributions of electrons emitted as a result of AMPS from SLG and clean Cu. Measurements were made on a SLG sample grown on polycrystalline Cu and on the underlying Cu surface after removing the SLG by argon ion sputtering. The measured AMPS spectra have been successfully reproduced using a model which consists of the weighted partial density of states (DOS), the positron kinetic energy distribution, and estimates of the electron escape probabilities. The weights can be rationalized using Auger Matrix elements \cite{Appelbaum1975}. Our results show that the surface DOS is directly reflected in the AMPS spectrum and demonstrate that AMPS is a top-most atomic layer selective probe of the electronic structure of fragile two-dimensional surfaces. 

The measurements were performed using a positron beam system equipped with a magnetic bottle time-of-flight (ToF) spectrometer. The details of the experimental setup are provided in supplemental material \cite{suppmat} and Ref. \cite{Mukherjee2016}. The presented data were collected using a 4 mCi $^{22}$Na source with $\sim$ 50-100 positrons per second reaching the sample resulting in measurement times of 12-48 hours. The kinetic energies of the AMPS electrons are attained from the electron ToF. The electron ToF is the time difference between the detection of the annihilation gamma photon and the detection of the electron. The magnetic bottle ToF spectrometer permits the collection of electrons ejected over $2\pi$ sr, and thus all data presented here are angle-integrated.

\begin{figure}
\centering
   \includegraphics[width=0.6\textwidth,keepaspectratio]{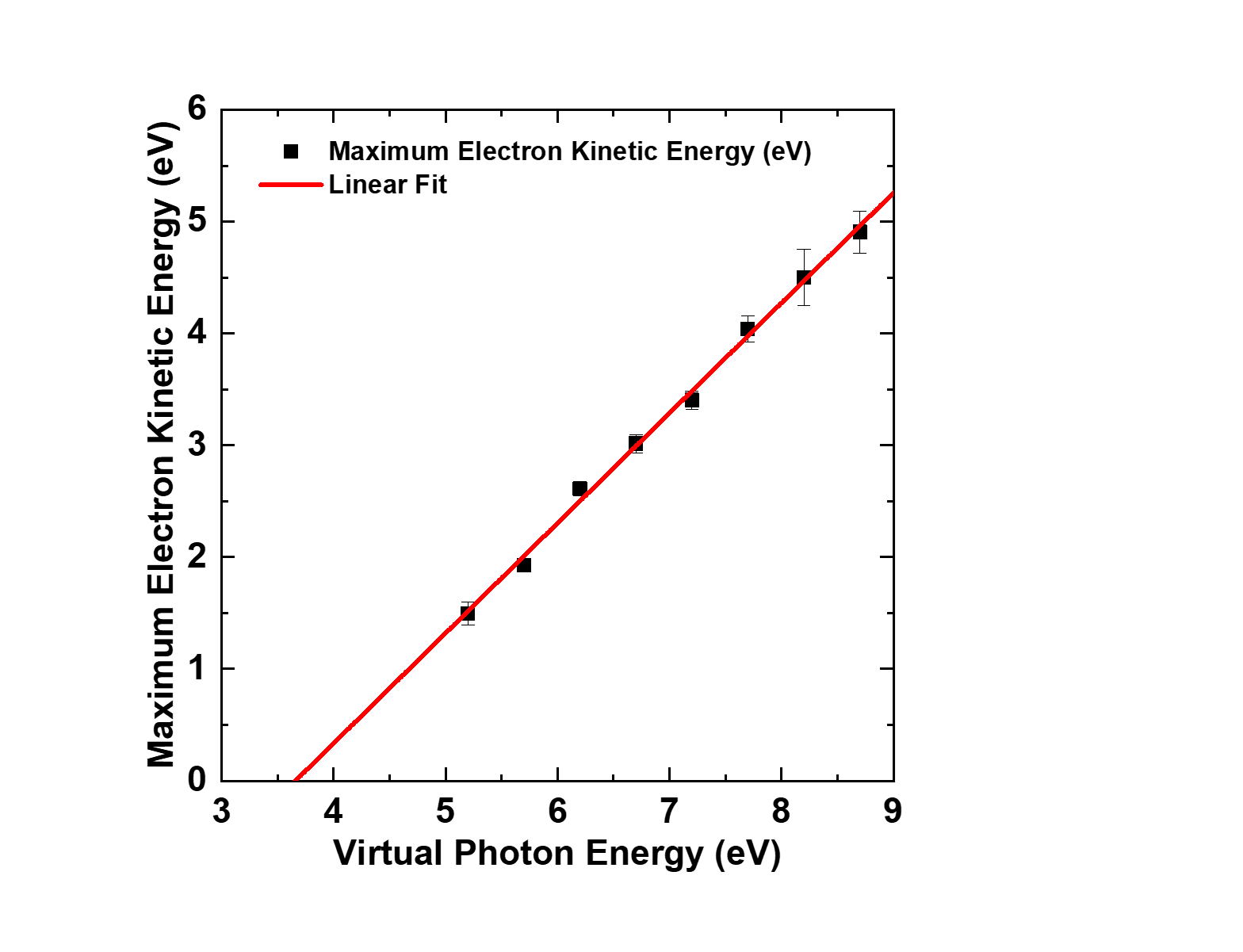}
    \caption{The maximum energies of the AMPS electrons from Cu, $KE^-_{max}$, as a function of the maximum virtual photon energies.}
    \label{fig:maxke}
\end{figure}

We show in Fig. \ref{fig:maxke} the variation of the maximum kinetic energy of electrons emitted via AMPS with the maximum virtual photon energy, $E_{vp,max}$, given by: 
\begin{equation}
    E_{vp,max} = KE^+_{max} + \varepsilon_{ss},
    \label{eq:vp}
\end{equation}
where $KE^+_{max}$ is the maximum positron kinetic energy and $\varepsilon_{ss}$ is the positron surface state binding energy. The maximum electron kinetic energy was determined after subtracting a background which consists solely of positron annihilation induced Auger electron spectra (PAES) \cite{Weiss1988}. PAES was obtained using measurements with incident positron beam kinetic energies less than 1.25 eV (see supplemental material for the unsubtracted AMPS and PAES data \cite{suppmat}). At these incident positron kinetic energies, only Auger electron emission is energetically possible \cite{Mukherjee2010}. After subtracting the PAES contributions, the maximum electron kinetic energy was determined from a straight line fit to the high energy edge of each AMPS spectrum. The variation of the maximum electron kinetic energies with $E_{vp,max}$ was fit with a straight line with a slope of $0.98 \pm 0.02$. This linear relationship between the maximum electron kinetic energy and the excitation energy was famously demonstrated in the photoelectric experiments of Millikan \cite{Millikan1916} and demonstrates that AMPS is a photoemission process as depicted in Fig. \ref{fig:amps}. A slope of unity implies that the maximum electron kinetic energy changes in accordance with the maximum positron kinetic energy. This means that the total energy of the incident positron is transferred to only one electron in the material and that measurable electron emission occurs only when the positron transitions to the positron ground state energy level of the surface state. An estimate of the energy levels of the positron in the surface state shows that for Cu the first excited state of positrons in the image potential-induced well is too shallow to permit electron emission while for SLG the first excited state does not exist. Therefore, we take the kinetic energy of an electron emitted as a result of AMPS to be: $KE^- = E_{vp} - \varepsilon_1 - \phi^-$, where $\varepsilon_1$ is the electron binding energy and $\phi^-$ is the electron work function.

\begin{figure}
    \centering
   \includegraphics[width=0.5\textwidth]{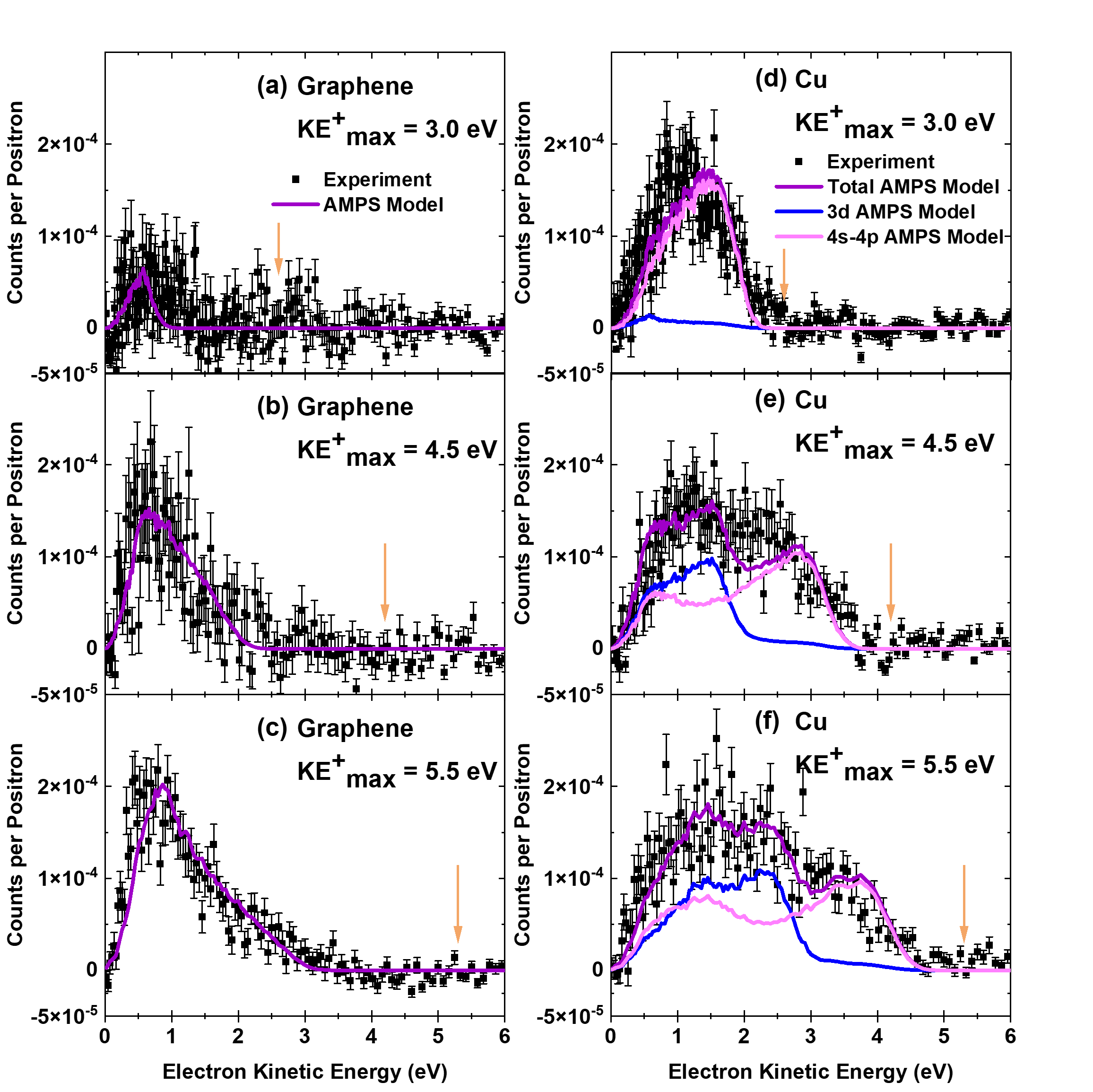}
    \caption{\small{The measured AMPS spectra in black for maximum incident positron beam energies of 3.0 eV, 4.5 eV, and 5.5 eV shown alongside a modelled AMPS spectra in violet. Panels (a)-(c) are the results for SLG. Panels (d)-(f) are the results for Cu where the Cu AMPS spectra consists of a weighted sum of 4$s$-$p$ (pink) and 3$d$ (blue) partial AMPS spectra as described in the text. The kinetic energy which corresponds to the Fermi Level is indicated by an orange arrow in each panel.}}
    \label{fig:fits}
\end{figure}

Figure \ref{fig:fits} shows the AMPS spectra (black squares) for SLG and Cu, after background subtraction, alongside a modelled AMPS spectra (violet solid lines) for maximum incident positron beam energies of 3.0 eV, 4.5 eV, and 5.5 eV. For both SLG and Cu, as the incident positron beam kinetic energy is increased from 1.25 eV, a low-energy AMPS peak emerges and continues to grow extending to higher maximum electron kinetic energies. The growth in intensity of the AMPS peak is due to the increased number of valence band states that can be excited due to the increased virtual photon energy. The intensities of the AMPS peaks are at least two orders of magnitude larger than what would be expected if positron sticking resulted in the emission of a ``real" photon followed by photoelectron emission \cite{Zhou2021} providing evidence for the virtual photon mediated electron emission process shown in Fig.\ref{fig:amps}. Additionally, the considerable differences in the overall shape, the peak intensity, and the maximum electron kinetic energies of the AMPS peaks from SLG on Cu and clean Cu obtained by removing just one atomic layer of carbon demonstrates the selectivity of AMPS to the top-most atomic layer.  

The AMPS electron photo-current as function of energy, $I(KE^-)$, has been modelled using: 
\begin{equation}
  I(KE^-)= P_{e}(KE^-)\ \int_{-\infty}^{\infty} 
  F(E_{vp}) \ d{E_{vp}}
  \int_{-\infty}^{E_F} 
  D_w(\varepsilon_1)  \ \delta(KE^-+\phi^{-}+\varepsilon_1-E_{vp}) \
  d\varepsilon. \\
  \label{eq:intensity}
\end{equation}
Here, $KE^-$ is the kinetic energy of the emitted AMPS electron, $P_{e}(KE^-)$ is the electron escape probability which weights the spectrum according to the direction of emission of the electron \cite{Hagstrum1966}, $F(E_{vp})$ is the virtual photon energy distribution, $E_F$ is the Fermi energy and $D_w(\epsilon_1)$ is the effective surface DOS probed by the virtual photon emitted following positron sticking. The virtual photon energy is given by equation \ref{eq:vp} with the added energy from $\phi_{c}$, the contact potential between the sample and spectrometer. Energy conservation is maintained through the Dirac delta function. If the positron beam is monochromatic, i.e., if $F(E_{vp})=\delta(E_{vp}-{\bar E_{vp}})$, then the photo-current $I(KE^-)$ reduces to $I(KE^-) = P_{e}(KE^-) D_w({KE^-+\phi^{-}-\bar E_{vp}}).$ The escape probability function $P_{e}(KE^-)$ rises fast from zero quickly reaching the asymptotic value of 0.5 and has appreciable influence on the shape of AMPS spectra only at lowest electron energies \cite{Hagstrum1966}. Hence, $I(KE^-)$ is directly reflective of the surface DOS $D_w(\varepsilon)$ and thus, our model is similar to that used for photoemission. The kinetic energy distributions of AMPS electrons calculated using Eq. \ref{eq:intensity} were used as an input to a SIMION 8.1 \cite{DAHL20003} simulation of our spectrometer to account for instrumental broadening of the outgoing electron energy distributions. The effect of the spectrometer response function is to smooth and slightly broaden the input kinetic energy distributions but the spectroscopic features of the input spectrum are maintained \cite{Chirayath2017,Fairchild2017}. Finally, an overall scale factor determined using a one-parameter least squares fit was applied to bring the experimental and calculated peaks into agreement.

For SLG (Fig. \ref{fig:fits}(a)-(c)), excellent agreement with experiment was found taking the effective surface DOS, $D_w(\epsilon)$, to be the total DOS of graphene calculated as described in \cite{Chirayath2017, Callewaert2018} (see supplemental material \cite{suppmat}). Through our modelling, we found that the surface state binding energy for SLG was 1.7 eV. There are currently no experimental measurements of the surface state binding energy of graphene/graphite in the literature \cite{Sferlazzo1988}. However, the surface state binding energy of SLG determined from our experiments is close to the binding energy of positrons on a diamond surface obtained using ab-initio calculations \cite{Callewaert2018}. 

The AMPS spectra from Cu (Fig. \ref{fig:fits}(d)-(f)) contains features (the step at $\sim$3 eV in Fig. \ref{fig:fits}(f) and at $\sim$2 eV in Fig. \ref{fig:fits}(e)) corresponding to the 3$d$ bands that are considerably suppressed in comparison to the photoemission spectra of Cu \cite{Rowe1974, Courths1984}. This implies that the bulk total DOS does not accurately represent $D_w(\epsilon)$ in Eq. \ref{eq:intensity}. Therefore, the Cu AMPS spectra were modelled using a weighted sum of partial AMPS spectra involving either 3$d$ or 4$s$-$p$ states (solid blue and pink lines respectively in Fig. \ref{fig:fits}(d-f)). These partial AMPS spectra were calculated using the partial DOS of bulk Cu \cite{Papa} which were shown to be in reasonable agreement with photoemission spectra of copper \cite{Courths1984}. Other positron-induced experiments, which have similar Auger matrix elements to AMPS, have also been described well using a weighted bulk DOS \cite{Mills1995}. The weights of the individual 3$d$ and 4$s$-$p$ spectra were determined using a least squares fit to the data collected with a maximum incident positron beam kinetic energy of 5.5 eV (Fig. \ref{fig:fits}(f)), since at this positron energy the greatest number of electronic states is probed while avoiding the influence of impact-induced secondary electron emission observed at higher positron energies \cite{Mukherjee2010}. In fitting the spectra of Fig. \ref{fig:fits}(d-e), these weighting factors were keep constant and only an overall scale factor determined from a one-parameter least squares fit to the data was used. A surface state binding energy of 3.0 eV was obtained through our fitting which is consistent with previous measurements \cite{Mukherjee2010,Mills1979}. 

The ratio of the weight of Cu 4$s$-$p$ bands to the 3$d$ bands is $24 \pm 1.6$ which shows that the contribution of the Cu 3$d$ bands to the AMPS spectra is suppressed in comparison to the $s$-$p$ bands on the surface. Similar effect has been observed in ion-neutralization spectroscopy (INS) \cite{Hagstrum1967, Hood1985} and the positronium time-of-flight spectroscopy of $d$-band metal surfaces \cite{Howell1987}. The reduced contribution of 3$d$ bands to ion neutralization spectra of $d$-band metals like Ni and Cu was understood in terms of the localized nature of 3$d$ orbital in comparison to the diffuse nature of the 4$s$-$p$ orbitals \cite{Hagstrum1967, Appelbaum1975, Hood1985}. Though significant differences exist between AMPS and ion neutralization, we can understand the AMPS spectra in similar terms - i.e. in terms of the DOS and the spatial extent of the orbitals forming the bands. In Cu, since the $s$-$p$ orbitals extend further into the vacuum, the electrons from these states have a greater probability of coupling to the range-limited virtual photon emitted following the sticking of positron a few angstroms outside the top-most atomic layer in comparison to the electrons from the highly localized 3$d$ states resulting in the enhancement of the $s$-$p$ states observed in the AMPS spectra of Cu. We note that our model successfully describes the AMPS line shape of both Cu and SLG without explicitly considering inelastic scattering of the outgoing electrons. This is reasonable given that the low-energy AMPS electrons, which are generated at the top-most atomic layer, have large IMFPs.

The surface selectivity and the shape of the kinetic energy distributions of the AMPS electrons, as described by equation \ref{eq:intensity}, can be understood in terms of an Auger matrix element, $M_{f,i}$, describing the positron sticking. The index $i$ represents the initial state in which the positron is in a scattering state and the electron is in the solid. In the final state $f$, the positron is in a bound surface state while the electron is ejected with a kinetic energy $KE^-$. This matrix element can include correlation effects via the electron-positron contact term factor, $\gamma_{i}$, \cite{Barbiellini1997} and can be written in a simple form like the Auger matrix element in Ref. \cite{Mills1983}:

\begin{equation}
 M_{f,i}=\langle f|W|i \rangle= \sqrt{\gamma_{i}}\int G(\bm{x})\psi^-_i(\bm{x})d^3\bm{x}.
\label{eq:mills}
\end{equation} 
Here, $W(\bm{x}-\bm{X})=\frac{e^{-\mu|\bm{x}-\bm{X}|}}{|\bm{x}-\bm{X}|}$,  $\psi^-_i(\bm{x})$ is the initial electron wavefunction, and $G(\bm{x})$ is a function that contains all the information of the matrix element, namely the screened interaction potential, the positron wavefunctions before and after sticking, and the electron wavefunction after emission integrated over the positron position coordinates, $\bm{X}$. Since the screened Coulomb potential limits the range of interaction to the Thomas-Fermi screening length $\frac{1}{\mu}$ $\sim$ 1 {\AA}, and since the trapped positron wavefunction has a limited spatial extent in the direction perpendicular to the surface \cite{Chirayath2017}, the function $G(\bm{x})$ selects only electron wavefunctions, $\psi^-_i(\bm{x})$, that have appreciable presence within a $\sim$ 1 {\AA} slab on the vacuum side of the top-most atomic layer. 

Since Eq. \ref{eq:intensity} derives from Fermi's Golden rule, we have:

\begin{equation}
   D_w(\epsilon)=\sum_i |M_{f,i}|^2 \delta(\epsilon -E_i),
\end {equation} 
where $E_i$ is the initial energy of the electron. $D_w(\epsilon)$ is therefore, electron DOS sampled by the virtual photon ($|M_{f,i}|^2$) at the solid surface. If $|M_{f,i}|^2$ is a constant, then $D_w(\epsilon)$ becomes proportional to the total DOS $D(\epsilon)$ as in the case for SLG. However, for systems like Cu with 3$d$ bands, $|M_{f,i}|^2$ strongly depends on the ejected electronic states \cite{Hood1985}. A rough estimate of $|M_{f,i}|^2$, obtained from a partial-wave analysis \cite{Mills1995}, gives an angular momentum $l$ dependence such as:

\begin{equation}
 |M_{f,i}|^2 \sim \gamma_{l} \frac{(kr)^{(2l+1)}}{(2l+1)}.
 \label{eq:kr}
\end{equation} 
For $l=0$, $|M_{f,i}|^2 \sim \gamma_{0}kr$. Here, $k$ is the wavevector of the electron, which is of the order of the Fermi wavevector $k_F$, and $r$ is the range over which the interaction occurs which is of the order of the Thomas-Fermi screening length. Using the Cu Wigner-Seitz radius $r_s=2.67$ au \cite{ashcroft1976} to calculate the Fermi wave vector, $k_F=1.92/r_s$, and the 
Thomas-Fermi screening length, $\frac{1}{\mu} =r=\sqrt{r_s}/1.56$ \cite{Barbiellini1989}, we obtain an estimate of ratio of the 4$s$-$p$ to the 3$d$ matrix elements to be about $23$ if we use the state dependent enhancement factors by Barbiellini et al., \cite{Barbiellini1997}. If we use the enhancement factors obtained after the phenomenological correction by Laverock et al. \cite{Laverock2010} we obtain a ratio of $19$ in excellent agreement with our measurement \footnote{Since state dependent enhancement factors were not available for copper in Laverock et al., we took the values reported for chromium.}. The estimate given in Eq. \ref{eq:kr} is consistent with the explanation of the suppression of the contribution of 3$d$ band in the positronium time-of-flight spectra from Ni \cite{Howell1987, Mills1995}.  

Our results provide key insights into both positron physics and the larger area of near-field surface probes. Our modelling provides an efficient way of determining the positron surface state binding energy of technologically relevant materials like graphene \cite{Mills1979,Sferlazzo1988}. Additionally, our results indicating the relevance of matrix element effects are key for the interpretation of similar techniques that have equivalent Auger matrix elements including: positronium time-of-flight spectroscopy \cite{Jones2016, Maekawa2021, Mills1983}, ion neutralization at metal surfaces \cite{Hagstrum1966, Appelbaum1975, Hood1985, Mudhafer2019}, the healing mechanism of excited molecules near metallic surfaces \cite{Barbiellini2006}, interatomic and intermolecular Coulombic decay \cite{jahnke2020}, and energy transfer in photonics \cite{King2012}.

AMPS as a surface spectroscopy has significant differences in comparison to related techniques like INS or positronium time-of-flight spectroscopy. By controlling the incident positron beam kinetic energy we can selectively probe regions of the DOS. For instance, we have explored regions of the DOS of Cu very near the Fermi Level with significant 4$s$-$p$ contributions and deeper regions that have more pronounced 3$d$ contributions. This energy control is missing in INS because the spectrum is independent of the incident particle energy. INS involves the removal of two electrons whereas in AMPS only one electron is ejected from the surface. Therefore, the AMPS spectra samples the surface DOS directly, more similar to PES, whereas the INS spectra reflects the self convolution of the surface DOS. Hence, AMPS is expected to be more sensitive to chemical changes at the surface \cite{Yang2019}. Moreover, the positron in AMPS never enters the solid and stays on the vacuum side of the sample. Positronium time of flight spectroscopy, on the other hand, requires that the positron be deposited deep enough so that non-thermal positronium formation is avoided. Thus, one can use low-energies and low positron-fluxes making AMPS an ideal probe of fragile 2D materials.  

Probing the surface DOS, near the Fermi Level, of exclusively the top-most atomic layer (without any contribution from the underlying substrate) makes AMPS a complementary technique to existing photoemission spectroscopies of 2D materials. A clear advantage of AMPS is the elimination of the non-trivial, secondary electron background which can influence PES analysis. In principle, a spectroscopy of positron-induced electrons performed using an electrostatic positron beam, with an angle-resolved detector or a Mott polarimeter, can provide the momentum or spin-resolved surface electronic DOS. Furthermore, one can improve the energy resolution of our technique to be comparable to PES by utilizing current advancements in the creation of intense, monoenergetic positron beams \cite{Marler2005,Mills2018,Golge2014,Hyodo2018}. Lastly, the possibility now exists of combining the present technique with positron diffraction to selectively measure both the atomic and the electronic structure of the top-most atomic layer of the sample surface \cite{Hyodo2018}.

\section*{Acknowledgments}
AHW was supported by the Welch Foundation grant No. Y-1968-20180324 and the NSF grants CHE 2204230, DMR 1508719, and DMR 1338130. BB was supported by the Ministry of Education and Culture (Finland) and acknowledges CSC-IT Center for Science, Finland, for computational resources.

A.J.F., V.A.C., A.H.W. designed the experiments and developed the theoretical model of the AMPS spectra. B.B estimated the AMPS matrix element. A.J.F, V.A.C, and B.B wrote the manuscript. R.W.G., and A.R.K. assisted in the interpretation and preparation of the manuscript.

\nocite{Mukherjee2011}
\nocite{Chirayath2019}
\nocite{Blochl1994}
\nocite{Kresse1996}
\nocite{KRESSE199615}
\nocite{Kresse1999}

\bibliographystyle{unsrt}  
\bibliography{main}

\end{document}